\documentclass[pdflatex,sn-mathphys-num]{sn-jnl}

\usepackage{graphicx}%
\usepackage{multirow}%
\usepackage{amsmath,amssymb,amsfonts}%
\usepackage{amsthm}%
\usepackage{mathrsfs}%
\usepackage[title]{appendix}%
\usepackage{xcolor}%
\usepackage{textcomp}%
\usepackage{manyfoot}%
\usepackage{booktabs}%
\usepackage{algorithm}%
\usepackage{algorithmicx}%
\usepackage{algpseudocode}%
\usepackage{listings}%

\theoremstyle{thmstyleone}%
\theoremstyle{thmstyletwo}%

\theoremstyle{thmstylethree}%

\begin{document}

\title[Domain-Adaptive Data Assimilation]{Domain-Adaptive Data Assimilation for Global AI Weather Forecasting}

\author[1]{\fnm{Minseok} \sur{Seo}}\email{minseok.seo@kaist.ac.kr}
\author[2]{\fnm{Noah} \sur{Brenowitz}}\email{nbrenowitz@nvidia.com}

\author[1]{\fnm{Doyi} \sur{Kim}}\email{doyi.kim@kaist.ac.kr}

\author[3]{\fnm{Hyesook} \sur{Lee}}\email{hyesook.lee@korea.ac.kr}

\author[1]{\fnm{Changick} \sur{Kim}}\email{changick@kaist.ac.kr}

\affil[1]{\orgdiv{Korea Advanced Institute of Science and Technology (KAIST)}}
\affil[2]{\orgdiv{NVIDIA}}
\affil[3]{\orgdiv{National Institute of Meteorological Sciences (NIMS)}}

\abstract{
AI weather forecasting models are commonly trained on the ERA5 reanalysis, which is unavailable in real time.
Operational deployment therefore relies on initial conditions produced by numerical or AI analysis systems that differ from those encountered during training.
This mismatch can degrade forecast skill, while retraining for every analysis system is costly.
Here, we present Domain-Adaptive Data Assimilation (DADA), an observation-guided framework that adapts external analyses to pretrained AI weather models.
Starting from a background state, DADA optimizes only an initial-state perturbation while keeping the forecast model frozen.
The perturbed state is propagated through the model, and its short-range trajectory is constrained by real-world observations through a learned observation operator.
The resulting initial condition is shaped jointly by observational constraints and the dynamics learned by the target model.
We evaluate DADA across five global AI weather models using backgrounds from the Global Forecast System and the AI-based HealDA.
Across deterministic and probabilistic forecasts, DADA substantially reduces short-range skill loss caused by changes in the initial-condition source.
More broadly, DADA turns observations into a common interface between independently developed analysis and forecasting systems, enabling pretrained AI weather models to accommodate evolving operational initial conditions without reconstructing ERA5 or retraining the forecast model.
}

\keywords{Domain Adaptation, Data assimilation, Weather forecasting, Climate, AI}

\maketitle

\section{Introduction}\label{sec1}
Recent advances in artificial intelligence (AI) have enabled global weather forecasting models to match state-of-the-art numerical weather prediction (NWP) systems and even outperform medium-range forecasting tasks~\cite{bi2023accurate,lam2023learning,price2025probabilistic,kochkov2024neural,bonev2025fourcastnet,kossaifi2026demystifying,pathak2022fourcastnet,bodnar2025foundation}.
These advances have relied heavily on reanalysis data accumulated systematically over several decades. 
In particular, most AI-based weather forecasting models are trained on ERA5, which provides hourly estimates of the global atmospheric state from 1940 to the present~\cite{hersbach2020era5, soci2024era5}. 
This record provides a consistent foundation for learning a wide range of atmospheric states and their underlying physical and dynamical relationships.

Given the current atmospheric state as an initial condition, these models autoregressively predict its future evolution.
However, ERA5~\cite{soci2024era5} is not available in real time: the preliminary ERA5T product is released approximately five days behind real time, while the quality-controlled final ERA5 product has a latency of approximately two to three months.

In operational settings, ERA5-trained forecasting models must therefore rely on initial conditions available at forecast time.
These initial conditions are commonly provided by operational NWP analyses, such as the Integrated Forecasting System (IFS)~\cite{ecmwf2026ifs} and the Global Forecast System (GFS)~\cite{noaa_gfs}, while recent AI-based analysis systems~\cite{xiao2023fengwu,li2024fuxi,zhu2026aligning, vonich2026atmospheric,chen2023towards,xiang2025adaf,xu2025fuxi,wang2024dabench,wang2025four,gupta2026healda, allen2025end,alexe2024graphdop,zhao2024omg,pinnington2026aifs} provide alternative ways of generating atmospheric states from observations and background information.

Despite their high quality, operational NWP analyses are produced using data assimilation systems~\cite{courtier1994strategy,le1986variational,bannister2017review,kalnay2003atmospheric}, numerical models, observation-processing procedures, and spatial resolutions that differ from those used to generate ERA5.
These differences can lead to shifts in variable-wise means and variances, spatial spectra, cross-variable balances, and error structures.
AI-based analysis systems provide an alternative source of operational initial conditions, but many are trained to reconstruct a prescribed reanalysis such as ERA5.
Because this reconstruction is inevitably imperfect, residual errors can remain in the generated analyses, leaving a distributional gap between these initial conditions and the ERA5 states used to train downstream forecasting models.
Thus, whether initial conditions are obtained from NWP-based or AI-based analysis systems, a source-dependent mismatch with the training distribution of pretrained forecasting models can persist.
Because AI forecasting models are optimized for the distribution of atmospheric states encountered during training, even modest deviations from this distribution in previously unseen initial states can degrade forecast performance~\cite{szegedy2013intriguing, goodfellow2014explaining,hendrycks2019benchmarking}.
Small discrepancies at initialization may also be amplified through autoregressive rollout, causing models initialized from operational analyses to underperform relative to evaluations based on ERA5 initial conditions~\cite{moldovan2026aifs, subich2025efficient, sadeghi2025gfs, wang2025development}. 

This mismatch can become increasingly important when a fixed AI forecasting model is deployed over an extended period.
Operational NWP systems continuously evolve through updates to observing networks, data assimilation methods, and physical parameterizations~\cite{atlas1997atmospheric,bauer2015quiet,bluestein2022atmospheric,canton2021world}, while the climate itself also changes over time. 
In contrast, a pretrained AI forecasting model remains tied to the characteristics of the reanalysis distribution on which it was trained.
Consequently, the operational input distribution may progressively deviate from the one encountered during training, creating a persistent challenge for deploying pretrained AI weather models in real-time forecasting systems. 
\begin{figure}[t]
    \centering
    \includegraphics[width=\columnwidth]{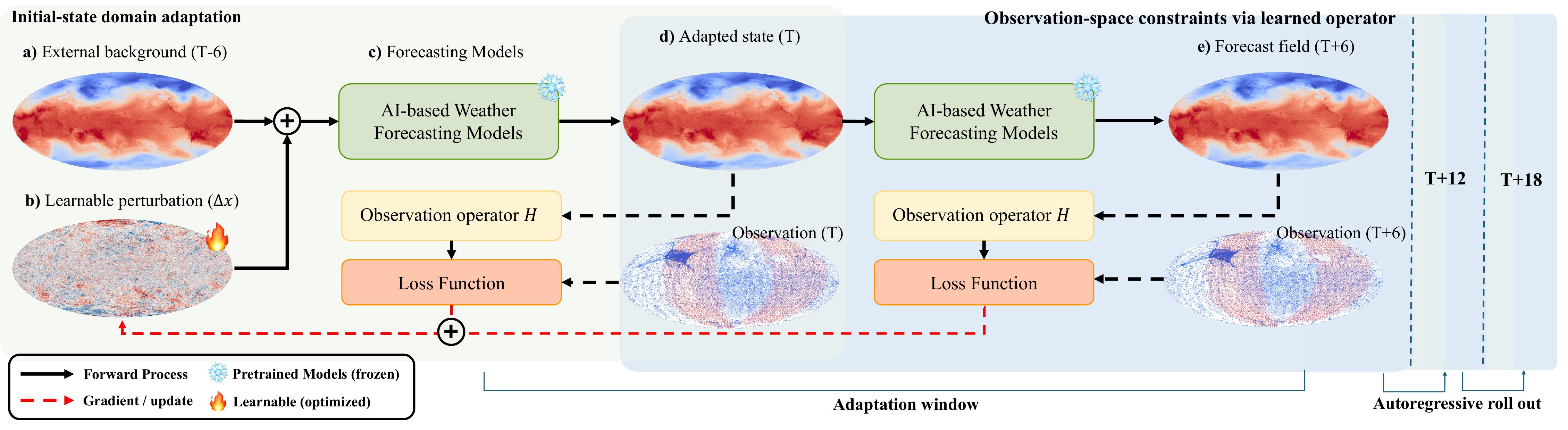}
\caption{
Overview of the proposed domain-adaptive data assimilation (DADA) method.
A learnable perturbation $\Delta\mathbf{x}$ is added to the background state at $T-6$, and the perturbed state is advanced using a pretrained, frozen AI weather forecasting model.
The resulting forecast state at $T$ is defined as the DADA analysis, $\mathbf{x}_{T}^{\mathrm{DADA}}$, and is mapped into observation space by a fixed neural observation operator $\mathcal{H}$.
The resulting observation-space estimate, $\hat{\mathbf{y}}_{T}=\mathcal{H}(\mathbf{x}_{T}^{\mathrm{DADA}})$, is compared with the real observations $\mathbf{y}_{T}$.
The frozen forecasting model is then applied autoregressively to produce the state at $T+6$, whose observation-space estimate, $\hat{\mathbf{y}}_{T+6}=\mathcal{H}(\hat{\mathbf{x}}_{T+6})$, is compared with $\mathbf{y}_{T+6}$.
The observation losses at $T$ and $T+6$ are combined and backpropagated through the frozen observation and forecasting models to update only $\Delta\mathbf{x}$.
Neither the forecasting model nor $\mathcal{H}$ is updated.
Instead, DADA estimates the perturbation through test-time optimization, producing an analysis and subsequent forecast trajectory that are more consistent with the observations.
}
    \label{fig:main}
\end{figure}

One possible solution is to retrain or fine-tune the forecasting model for each analysis system. However, training a separate model whenever the analysis system or input distribution changes is computationally expensive and difficult to maintain in continuously evolving operational environments~\cite{subich2025efficient}.
Directly using ERA5, which was used for model training, as an adaptation target is also impractical because ERA5 is not available at operational forecast time.
What is therefore needed is a source of information that can adapt initial conditions generated by different analysis systems to a pretrained forecasting model without directly targeting a particular analysis system or reanalysis.
Real-world observations can provide such a common adaptation signal.
Regardless of whether the background is generated by GFS, IFS, or an AI-based analysis system, observations provide information about the same underlying atmospheric state.
In particular, if the relationship between ERA5 states and real-world observations can be exploited, discrepancies in observation space can provide information about how an external analysis differs from the ERA5-based states encountered during model training.
Thus, observations do not directly provide ERA5 itself, but can serve as an adaptation signal for adjusting backgrounds from different sources toward states that are more compatible with an ERA5-trained forecasting model.

From this perspective, we introduce Domain-Adaptive Data Assimilation (DADA). As illustrated in Fig.~\ref{fig:main}, DADA is a data assimilation framework that adapts background states generated by different analysis systems to pretrained AI weather forecasting models. Starting from a given background, DADA optimizes only a perturbation to the initial state while keeping the forecast model parameters frozen. The perturbed state is propagated through the target forecasting model, and the resulting short-range forecast trajectory is compared with real-world observations through a learned observation operator that captures the relationship between ERA5 states and observations. By minimizing discrepancies in observation space, DADA adjusts the background toward a state that is more consistent with real-world observations while also being more compatible with the states and dynamics learned by the pretrained forecasting model. This procedure generates a case-specific initial condition without directly reconstructing a prescribed reanalysis target or retraining the forecasting model.

We evaluate DADA across five global AI weather forecasting models using backgrounds from the NWP-based GFS and the AI-based HealDA. Across deterministic and probabilistic forecasts, DADA consistently reduces the performance degradation caused by changes in the initial-condition source across models, variables, and lead times. These results show that, even without directly providing ERA5, real-world observations can serve as an effective signal for adapting initial conditions generated by different analysis systems to frozen AI forecasting models by exploiting the learned relationship between ERA5 states and observations.

\begin{figure}[t]
    \centering
    \includegraphics[width=\columnwidth]{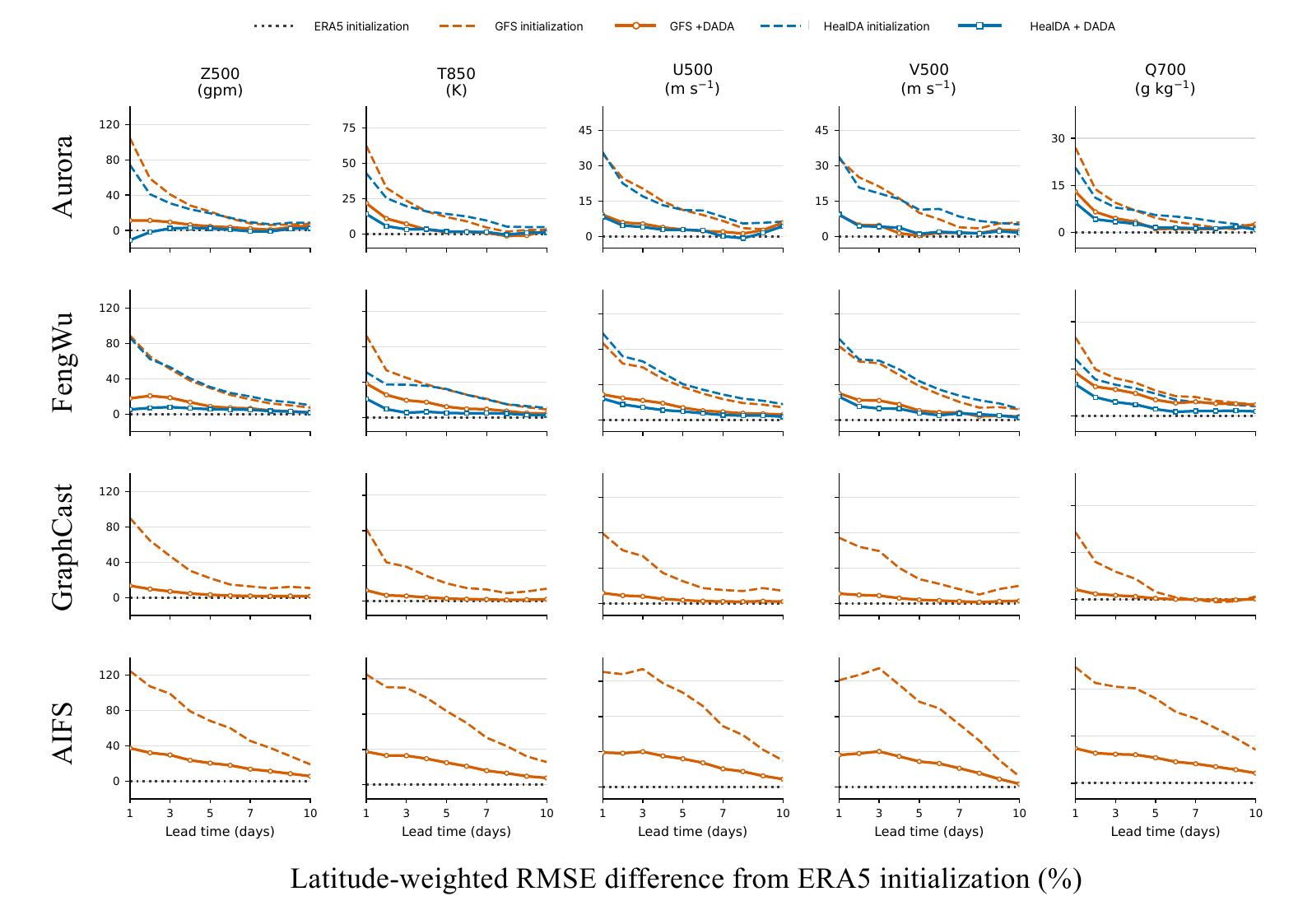}
    \caption{DADA adapts initial conditions across AI weather forecasting models. Percentage changes in latitude-weighted RMSE relative to ERA5 initialization are shown for 1 to 10-day forecasts from Aurora, FengWu, GraphCast and AIFS (rows) for Z500, T850, U500, V500 and Q700 (columns). All forecasts are verified against ERA5. Black dotted lines indicate ERA5 initialization; orange dashed and solid lines indicate GFS and GFS+DADA; and blue dashed and solid lines indicate HealDA and HealDA+DADA, respectively. Zero indicates the same RMSE as ERA5 initialization, whereas negative values indicate lower RMSE. Curves show the mean over 24 forecasts initialized at 06 UTC on the first and fifteenth days of each month in 2023. HealDA results are shown where available.}
    \label{fig:main_ex1}
\end{figure}

\section{Results}\label{sec2}
\subsection{Adaptation across forecasting models and initial conditions}
We first evaluated whether DADA can reduce the domain gap caused by changes in the initial-condition source across AI weather forecasting models.
We considered four deterministic forecasting models, Aurora, FengWu, GraphCast, and AIFS, and three background states, ERA5, operational GFS analysis, and HealDA analysis.

DADA uses a 12-h assimilation window from $T-6\,\mathrm{h}$ to $T+6\,\mathrm{h}$, with observation losses evaluated at $T$ and $T+6\,\mathrm{h}$.
The purpose of this experiment is not to benchmark DADA against existing data assimilation systems under strictly matched latency, but to quantify how much initialization mismatch can be reduced by adapting the same background to a pretrained forecasting model.
The sensitivity to assimilation-window length and the contribution of future observations are analyzed separately in supplementary material.

For each non-ERA5 background, we compared direct initialization from the original background with initialization from the DADA-adapted state, using ERA5 initialization as a reference.
Figure~\ref{fig:main_ex1} shows the percentage difference in latitude-weighted RMSE relative to ERA5 initialization.
Direct initialization from GFS or HealDA generally produced larger forecast errors than ERA5 initialization.
This does not imply that ERA5 is intrinsically more accurate, but instead reflects differences in compatibility with forecasting models trained primarily on ERA5.

DADA substantially reduced this degradation for both GFS and HealDA backgrounds.
Across forecasting models, variables, and lead times, forecasts initialized from DADA-adapted states were consistently closer to the ERA5-initialized reference than those initialized directly from the corresponding backgrounds.
These results show that DADA can adapt initial conditions from different analysis systems to frozen AI forecasting models and mitigate forecast degradation caused by changes in the initial-condition source.

\begin{figure}[t]
    \centering
    \includegraphics[width=\columnwidth]{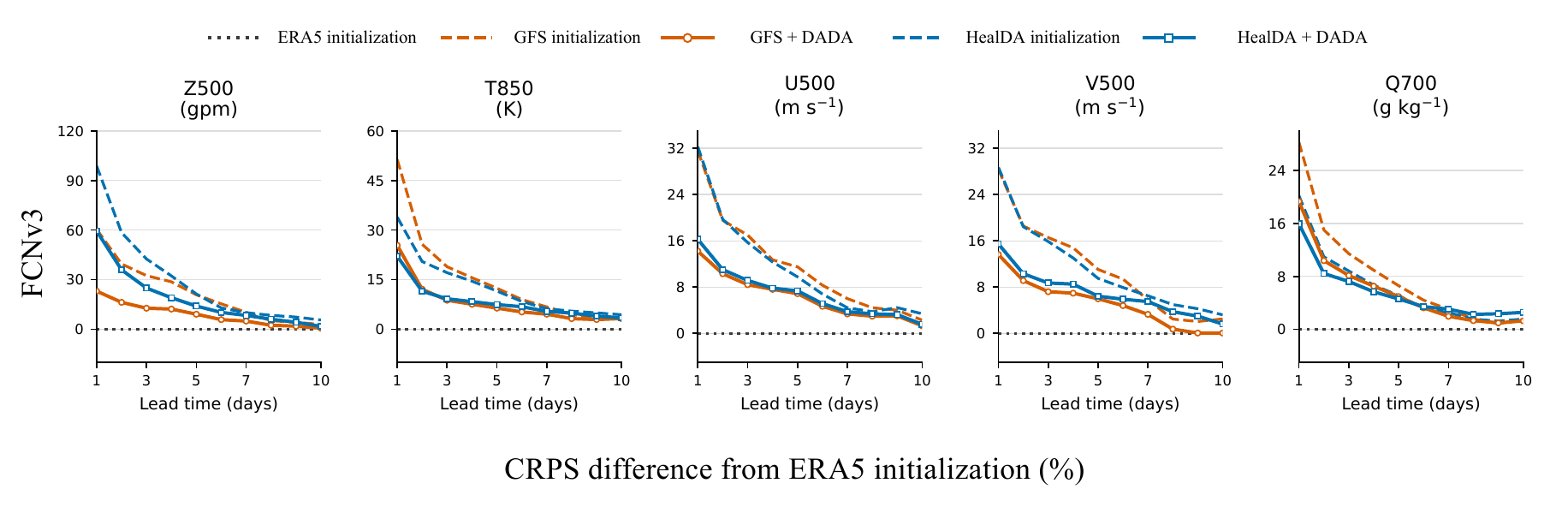}
    \caption{DADA improves probabilistic forecasts initialized from GFS and HealDA analyses. Percentage differences in the continuous ranked probability score (CRPS) relative to ERA5 initialization are shown for 10-member stochastic FCNv3 forecasts of Z500, T850, U500, V500 and Q700 at lead times of 1 to 10 days. Zero indicates the same CRPS as ERA5 initialization, while lower values indicate better probabilistic forecast skill. Black dotted lines denote ERA5 initialization. Orange dashed and solid lines denote GFS and GFS+DADA initialization, respectively, while blue dashed and solid lines denote HealDA and HealDA+DADA initialization, respectively. Each curve shows the mean over 24 forecast cases initialized at 06 UTC on the first and fifteenth days of each month in 2023. All forecasts were verified against ERA5.}
    \label{fig:main_ex2}
\end{figure}

\noindent\textbf{Probabilistic forecast skill with FCNv3.}
We further evaluated DADA using the 10-member stochastic FCNv3 ensemble.
As shown in Figure~\ref{fig:main_ex2}, forecasts initialized from GFS or HealDA generally exhibited higher CRPS~\cite{hersbach2000decomposition} than those initialized from ERA5, particularly at short lead times.
DADA substantially reduced this initialization gap across most variables, with the strongest and most persistent improvements observed for Z500.
Differences among the initialization sources gradually decreased with lead time, consistent with the declining influence of the initial conditions as forecast errors grow.
Together with the deterministic results, these findings show that DADA mitigates forecast degradation caused by initialization mismatch in both deterministic and probabilistic forecasting models.

\begin{figure}[t]
    \centering
    \includegraphics[width=\columnwidth]{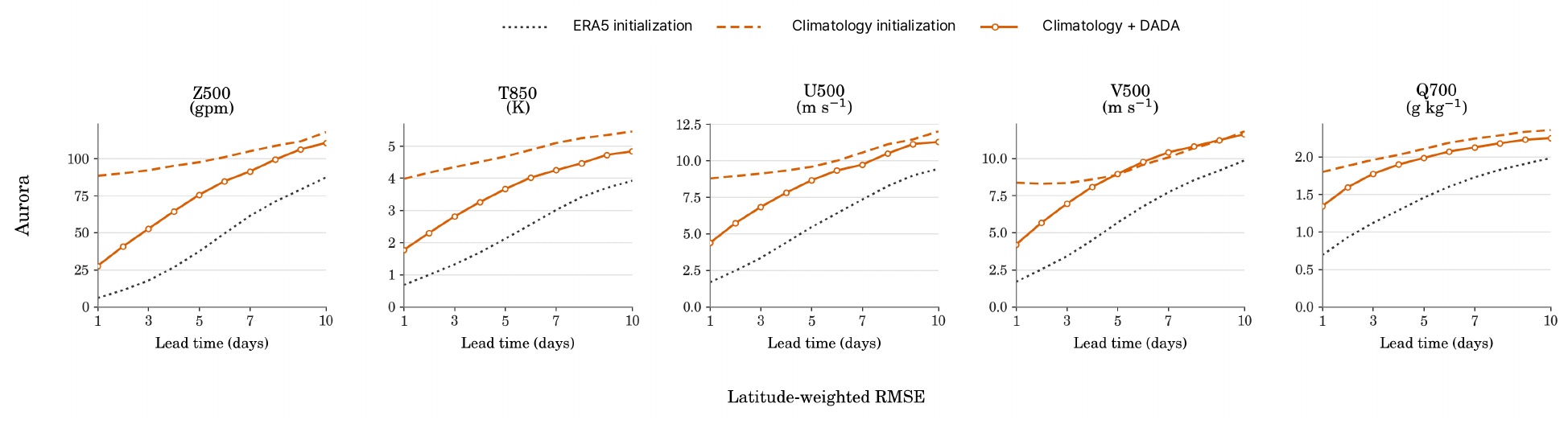}
    \caption{DADA recovers short-range forecast skill from a climatological background. Latitude-weighted RMSEs of Z500, T850, U500, V500 and Q700 are shown from Days 1 to 10 for Aurora initialized from ERA5, climatology and climatology adapted by DADA. Black dotted lines denote ERA5 initialization, orange dashed lines denote climatology initialization, and orange solid lines with open circles denote climatology+DADA initialization. Monthly climatological backgrounds were constructed from the 1990–2019 ERA5 climatology and matched to the calendar month and UTC cycle.}
    \label{fig:main_ex4}
\end{figure}

\subsection{Assimilation from a climatological background}

We tested DADA under an intentionally limited background by using the 1990--2019 ERA5 climatology as the background for Aurora.
Each climatological background was matched to the month and UTC cycle of the corresponding forecast case in 2023.
As shown in Figure~\ref{fig:main_ex4}, direct initialization from climatology produced large forecast errors because the background contains little case-specific synoptic information.
DADA substantially reduced the latitude-weighted RMSE across variables and lead times from Days 1 to 10, although it did not recover the performance of ERA5 initialization.
These results show that DADA can recover substantial forecast skill even from a background with limited information about the actual atmospheric state.
They also indicate that the achievable performance depends on both the information contained in the background and the dynamics represented by the frozen forecasting model.

\begin{figure}[t]
    \centering
    \includegraphics[width=\columnwidth]{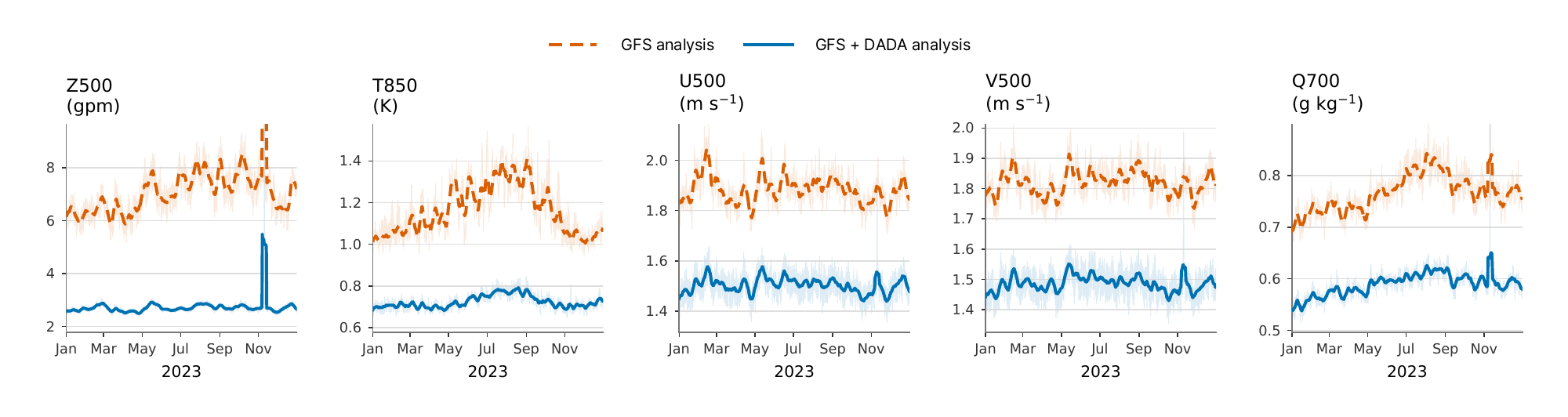}
    \caption{Adaptation of GFS analysis fields by DADA throughout 2023. Latitude-weighted RMSEs of Z500, T850, U500, V500, and Q700 relative to ERA5 are shown for the original GFS states and the corresponding DADA-adapted states generated every 6 h throughout 2023 using Aurora. ERA5 serves as a reference for the training regime of Aurora rather than for a direct comparison of data assimilation accuracy. Lines with reduced opacity show individual analysis times, and darker lines show 7-day moving averages. Orange dashed lines denote GFS, and blue solid lines denote GFS adapted by DADA.}
    \label{fig:main_ex5}
\end{figure}

\subsection{Adaptation of GFS analysis fields throughout 2023}

The previous experiments showed that DADA reduces the forecast degradation associated with GFS initialization.
We next examined how DADA modifies the original GFS background throughout 2023.
Using Aurora as the forecasting model, we generated DADA analyses every 6 h and measured the discrepancy of the states before and after adaptation from ERA5 using latitude-weighted RMSE.
ERA5 serves here as a reference for the primary training distribution of Aurora.

As shown in Figure~\ref{fig:main_ex5}, DADA-adapted states were consistently closer to ERA5 than the original GFS states across Z500, T850, U500, V500, and Q700, with this behavior remaining stable throughout the year.
This comparison is not intended to benchmark the analysis accuracy of DADA against GFS, but to examine whether DADA systematically modifies GFS backgrounds toward states that are more compatible with the training regime of the pretrained forecasting model.
Together with the forecast results, these findings are consistent with reduced initial-condition domain mismatch after adaptation.
The respective contributions of domain adaptation and short-range trajectory constraints to forecast performance are analyzed separately in supplementary material.

\begin{figure}[t]
    \centering
    \includegraphics[width=\columnwidth]{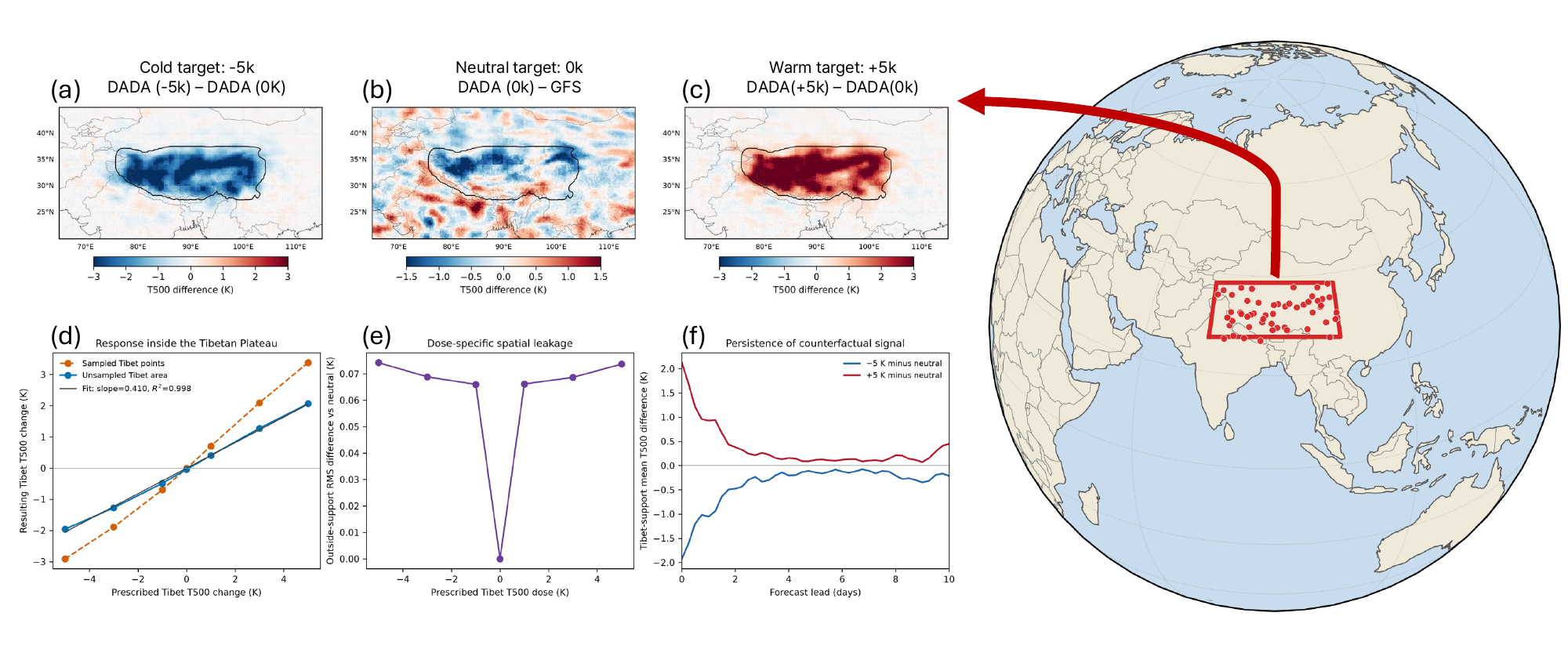}
    \caption{Local adjustment of DADA analyses through sparse observation targets. DADA analyses were generated from the same GFS background at 06 UTC on 15 March 2023 using a frozen Aurora model, while prescribing different temperature changes to sparse synthetic T500 observations over the Tibetan Plateau. (\textbf{a}, \textbf{c}) T500 differences between the analyses generated using $-5$ K and $+5$ K targets, respectively, and the analysis generated using the neutral target. (\textbf{b}) T500 difference between the neutral-target DADA analysis and the GFS control. Black contours delineate the Tibetan Plateau target region defined using elevation and spatial tapering. (\textbf{d}) Relationship between prescribed observation targets ranging from $-5$ K to $+5$ K and the mean T500 responses at sampled observation points and within the unsampled portion of the target region. The black line denotes a linear regression fitted to the response in the unsampled region. (\textbf{e}) T500 RMS differences outside the Tibetan Plateau target region between each target experiment and the neutral experiment. These values quantify the extent to which the response induced within the Tibetan Plateau spreads beyond the intended target region. (\textbf{f}) Evolution of the Tibetan Plateau mean T500 differences between the $\pm 5$ K target experiments and the neutral experiment over the 10-day forecasts. The map on the right shows the Tibetan Plateau experiment region and the sparse observation points used for DADA optimization.}
    \label{fig:main_ex6}
\end{figure}

\subsection{Influence of synthetic observation targets on DADA analyses}
We conducted a controlled experiment to examine whether DADA responds systematically to the direction and magnitude of observation targets.
All experiments used the same GFS background at 06 UTC on 15 March 2023 and the same frozen Aurora model.
ERA5 T500 values at sparse grid points over the Tibetan Plateau were used as the neutral target, and artificial temperature perturbations ranging from $-5$ K to $+5$ K were added at the same locations.

Figure~\ref{fig:main_ex6}(b) shows the difference between the neutral-target DADA analysis and the original GFS state.
When cold and warm perturbations were imposed, the resulting DADA analyses changed in opposite directions within the target region, as shown in Figure~\ref{fig:main_ex6}(a,c).
The sign of the analysis response was consistent with the prescribed temperature change.
Figure~\ref{fig:main_ex6}(d) further shows that the response increased approximately linearly with the target magnitude at both sampled and unsampled grid points within the target region, indicating that the influence of sparse observations extended to nearby unobserved states.
At the same time, the T500 RMS difference outside the target region remained below approximately $0.08$ K across the experiments (Figure~\ref{fig:main_ex6}(e)), showing that the response remained largely localized.
Finally, the differences induced by the cold and warm targets weakened during the subsequent 10-day forecasts but retained opposite signs and evolved into distinct forecast trajectories (Figure~\ref{fig:main_ex6}(f)).

These results show that DADA does not simply reconstruct a single fixed ERA5 analysis.
Even with the same background and forecasting model, DADA produces different analyses and forecast trajectories in response to different observation constraints.

\section{Discussion}
In this study, we introduced DADA, an observation-based domain adaptation framework that adapts background states from different analysis systems to frozen AI weather forecasting models.
DADA improved forecast performance across four deterministic models, Aurora, FengWu, GraphCast, and AIFS, as well as the probabilistic stochastic FCNv3 model.
For backgrounds from both the NWP-based GFS and the AI-based HealDA, DADA consistently reduced the forecast degradation associated with changes in the initial-condition source.
Across the year-long evaluation with GFS backgrounds, the DADA-adapted states also remained consistently closer to the ERA5-based training regime of Aurora than the original GFS states.
This comparison should not be interpreted as a direct benchmark of DADA against GFS as a data assimilation system, since the two systems use different observational information and assimilation windows.
Rather, it provides evidence that the forecast improvements are accompanied by systematic changes in the background toward states that are more compatible with the pretrained forecasting model.

The controlled synthetic-observation experiment further clarifies the role of observational constraints in DADA.
With the background and forecasting model fixed, changing only the observation targets produced systematic changes in the resulting analyses and subsequent forecast trajectories.
The direction and magnitude of the analysis response followed the prescribed observation perturbations, while the response remained largely localized to the target region.
These results indicate that DADA does not simply map different backgrounds toward a single fixed ERA5-like analysis.
Instead, the adapted state depends on both the observational constraints and the dynamics of the target forecasting model.

DADA does not aim to reconstruct a prescribed reanalysis target.
Instead, it optimizes a perturbation to the initial state and propagates the modified state through a frozen forecasting model, while constraining the resulting state and short-range trajectory in observation space.
The resulting analysis is therefore generated through the dynamics of the target forecasting model rather than by directly adding a correction to the background at the analysis time.
The supplementary experiments provide further insight into this behavior.
Direct supervision using sparse ERA5 states produced larger improvements than observation-space supervision, consistent with the benefit of reducing the mismatch between the background and the ERA5-based training distribution.
At the same time, extending the assimilation window further improved forecast performance, indicating an additional contribution from constraining the short-range forecast trajectory.
Together, these results are consistent with two complementary roles of DADA: adapting the initial state toward a regime that is more compatible with the pretrained forecasting model and constraining its subsequent trajectory using observations.

DADA should not, however, be interpreted as requiring no training.
When applied to a new background or forecast case, the parameters of the target forecasting model remain fixed, but the learned observation operator $H$ must be trained in advance.
Thus, DADA performs case-specific adaptation without retraining the forecasting model, while still relying on a pretrained observation operator.
DADA is also not a standalone analysis system that operates independently of existing data assimilation systems.
It requires a background state that already contains information about the atmospheric state and a forecasting model capable of propagating the adapted state.

The climatological-background experiment illustrates this dependence.
Although DADA substantially recovered forecast skill from a background containing little case-specific synoptic information, it did not reach the performance obtained with ERA5 initialization.
This result indicates that observation-based adaptation cannot fully recover atmospheric information that is absent from the background within a limited assimilation window.
The achievable analysis and forecast trajectories are also constrained by the range of states and dynamics represented by the frozen forecasting model.
DADA is therefore best understood not as a replacement for existing data assimilation systems, but as an observation-based adaptation layer that connects existing analysis systems with pretrained AI weather forecasting models.
Our results show that such adaptation can reduce the sensitivity of frozen AI forecasting models to changes in the initial-condition source without reconstructing a fixed reanalysis target or retraining the forecasting model.

\section{Methods}
\subsection{DADA framework}

The proposed Domain-Adaptive Data Assimilation (DADA) framework performs case-specific adaptation of the initial condition while keeping the target AI forecasting model and observation operator fixed.
Starting from an external background, DADA optimizes only an additive perturbation to the initial state.
Rather than reconstructing a prescribed reference analysis such as ERA5, DADA identifies a model-specific atmospheric state whose analysis and subsequent short-range forecast trajectory are consistent with real-world observations and the learned dynamics of the target forecasting model.

Let $M_{\theta}$ denote a pretrained and frozen AI weather forecasting model, and let $H_{\phi}$ denote a pretrained and frozen observation operator. Given a background state $\mathbf{x}^{b}_{t-6}$ at time $t-6$, DADA first normalizes the background using the channel-wise mean $\boldsymbol{\mu}$ and standard deviation $\boldsymbol{\sigma}$. An optimizable initial-state perturbation $\mathbf{p}_{t-6}$ is then added in the normalized state space. The perturbed initial state is defined as

\begin{equation}
\widetilde{\mathbf{x}}_{t-6}
=
\boldsymbol{\mu}
+
\boldsymbol{\sigma} \odot
\left(
\frac{\mathbf{x}^{b}_{t-6}-\boldsymbol{\mu}}
{\boldsymbol{\sigma}}
+
\mathbf{p}_{t-6}
\right),
\label{eq:DADA_perturbed_state}
\end{equation}

where $\odot$ denotes element-wise multiplication. The perturbation $\mathbf{p}_{t-6}$ spans all prognostic input channels of the target forecasting model and has the same spatial resolution as the background state. It is initialized to zero independently for each assimilation cycle.

The perturbed initial state is propagated to the analysis time through the frozen forecasting model:

\begin{equation}
\mathbf{x}^{\mathrm{DADA}}_{t}
=
M_{\theta}
\left(
\widetilde{\mathbf{x}}_{t-6}
\right).
\label{eq:DADA_analysis}
\end{equation}

We define $\mathbf{x}^{\mathrm{DADA}}_{t}$ as the DADA analysis. The analysis is therefore not obtained by directly adding a correction to the background state at time $t$. Instead, it is generated by propagating the perturbed state from time $t-6$ through the dynamics of the target forecasting model.

To constrain the subsequent forecast trajectory, the DADA analysis is propagated for an additional forecast step:

\begin{equation}
\widehat{\mathbf{x}}_{t+6}
=
M_{\theta}
\left(
\mathbf{x}^{\mathrm{DADA}}_{t}
\right).
\label{eq:DADA_forecast}
\end{equation}

The atmospheric states at times $t$ and $t+6$ are mapped into their corresponding observation spaces using the observation operator:

\begin{equation}
\widehat{\mathbf{y}}_{t}
=
H_{\phi,t}
\left(
\mathbf{x}^{\mathrm{DADA}}_{t}
\right),
\label{eq:DADA_obs_t}
\end{equation}

\begin{equation}
\widehat{\mathbf{y}}_{t+6}
=
H_{\phi,t+6}
\left(
\widehat{\mathbf{x}}_{t+6}
\right).
\label{eq:DADA_obs_t6}
\end{equation}

For each observation time $\tau \in \{t,t+6\}$, the observation loss is defined as a latitude-weighted and variable-normalized mean squared error:

\begin{equation}
\mathcal{L}_{\mathrm{obs},\tau}
=
\frac{
\displaystyle
\sum_{i \in \Omega_{\tau}}
w_i
\left(
\frac{
\widehat{y}_{\tau,i}
-
y^{\mathrm{obs}}_{\tau,i}
}{
s_{v(i)}
}
\right)^2
}{
\displaystyle
\sum_{i \in \Omega_{\tau}} w_i
},
\qquad
\tau \in \{t,t+6\}.
\label{eq:DADA_obs_loss}
\end{equation}

Here, $\Omega_{\tau}$ denotes the set of available observations at time $\tau$, $y^{\mathrm{obs}}_{\tau,i}$ is the corresponding real-world observation, and $\widehat{y}_{\tau,i}$ is the value predicted in observation space. The term $w_i$ is the latitude-dependent weight of observation $i$, and $s_{v(i)}$ is the normalization scale associated with its variable or channel.

The total DADA objective assigns equal weight to the observation losses at the analysis and subsequent forecast times:

\begin{equation}
\mathcal{L}_{\mathrm{DADA}}
=
\frac{1}{2}
\mathcal{L}_{\mathrm{obs},t}
+
\frac{1}{2}
\mathcal{L}_{\mathrm{obs},t+6}.
\label{eq:DADA_total_loss}
\end{equation}

For each assimilation cycle, only the initial-state perturbation $\mathbf{p}_{t-6}$ is optimized. The parameters of $M_{\theta}$ and $H_{\phi}$ remain fixed throughout the optimization. The perturbation is initialized to zero and optimized independently for each cycle using AdamW for 100 iterations, with a learning rate of $2 \times 10^{-3}$ and zero weight decay. After convergence, $\mathbf{x}^{\mathrm{DADA}}_{t}$ is retained as the final analysis and used to initialize the subsequent forecast. In this way, DADA adapts each background state to the observations and to the dynamics of the target forecasting model without retraining the forecasting model itself.
\subsection{Learned observation operator H}

The observation operator $H_{\phi}$ is the only component of DADA that requires offline training. It maps gridded atmospheric states to values corresponding to the locations, vertical coordinates, and types of real-world observations. Once trained, $H_{\phi}$ is frozen during the optimization of the DADA initial state. It provides a differentiable interface through which gradients of the observation-space loss are propagated through the forecasting model to the initial-state perturbation. Neither observed values nor observation innovations are included in the inputs to $H_{\phi}$. Real observations are used only as supervised targets during the training of $H_{\phi}$ and as targets of the observation-space loss during data assimilation.

For observation $i$, the predicted value in observation space is written as

\begin{equation}
\widehat{y}_{i}
=
H_{\phi}
\left(
\mathbf{x}_{i},
\boldsymbol{\eta}_{i}
\right),
\label{eq:h_definition}
\end{equation}

where $\mathbf{x}_{i}$ denotes the atmospheric state extracted at the observation location, and $\boldsymbol{\eta}_{i}$ denotes an observation-specific query containing information about the observation location, time, vertical coordinate, viewing geometry, observation type, and satellite. The output $\widehat{y}_{i}$ is a scalar prediction corresponding to observation $i$.

\subsubsection{Atmospheric-state and observation-query inputs}

The observation operator does not process the entire global grid as an image. Instead, the gridded atmospheric state is horizontally bilinearly interpolated to the latitude and longitude of each observation to construct a local atmospheric column, and each observation is predicted independently.

The atmospheric-state input is constructed from variables compatible with each target forecasting model. For example, the FCNv3-compatible input contains 72 atmospheric variables. The seven surface variables are 10-m and 100-m zonal and meridional winds, 2-m temperature, mean sea-level pressure, and total column water vapour. The vertical variables are zonal wind, meridional wind, geopotential, temperature, and specific humidity at 13 pressure levels: 50, 100, 150, 200, 250, 300, 400, 500, 600, 700, 850, 925, and 1000 hPa.

The observation query contains sine and cosine representations of latitude and longitude, the time offset between the assimilation-cycle centre and the observation time, observation pressure, observation height, station elevation, satellite and solar viewing geometry, model orography, land-sea mask, local orographic variability, and the elevation difference between the station and model terrain. Observation type, ATMS channel, and satellite identity are represented using learned embeddings.

Observations within each assimilation cycle are distributed within $\pm 3$ h of the cycle centre. The operator does not explicitly interpolate atmospheric states between adjacent times. Instead, it uses a single atmospheric state at the cycle centre and provides the difference between the observation time and the cycle time as an input query. Therefore, $H_{\phi}$ is a three-dimensional cycle-state observation operator with an observation-time-offset input, rather than a four-dimensional temporal-interpolation operator.

\subsubsection{Network architecture}

The observation operator consists of a surface-state encoder, a vertical-profile encoder, an observation-query encoder, a shared multilayer perceptron, and observation-type-specific output heads.

The seven surface variables are transformed into a 128-dimensional representation using a linear layer, layer normalization, and a GELU activation. For the vertical state, a six-dimensional token is constructed at each pressure level from zonal wind, meridional wind, geopotential, temperature, specific humidity, and normalized log pressure. The 13 vertical tokens are projected into 128-dimensional representations and processed by a two-layer Transformer encoder with four attention heads. Mean pooling over the pressure levels produces a 128-dimensional vertical-profile representation.

The continuous observation metadata and the embeddings of observation type, channel, and satellite identity are combined and transformed into a 96-dimensional representation by the query encoder. The 128-dimensional surface representation, 128-dimensional vertical-profile representation, and 96-dimensional query representation are concatenated into a 352-dimensional feature vector. A shared multilayer perceptron transforms this vector into a 64-dimensional feature representation. Separate output heads are used for ATMS brightness temperature, temperature, specific humidity, zonal wind, meridional wind, station pressure, and GNSS-RO bending angle. ATMS channels 2--22 share a common ATMS output head and are distinguished by learned channel embeddings, allowing the operator to model channel-dependent observation mappings within a shared architecture.

\subsubsection{Physical interpolation and learned residuals}

For observations that can be related directly to model variables through physical interpolation, the neural network is not required to learn the complete observation mapping from scratch. Temperature, specific humidity, zonal wind, and meridional wind are first horizontally interpolated to the observation location and then vertically interpolated to the observation pressure using log-pressure interpolation. A physical baseline for station pressure is calculated using mean sea-level pressure, 2-m temperature, station elevation, model orography, and the hypsometric relation.

For these observations, $H_{\phi}$ learns the residual between the physical baseline and the real observation. The final observation-space prediction is written as

\begin{equation}
H_{\phi}
\left(
\mathbf{x}_{i},
\boldsymbol{\eta}_{i}
\right)
=
H_{0}
\left(
\mathbf{x}_{i},
\boldsymbol{\eta}_{i}
\right)
+
\Delta H_{\phi}
\left(
\mathbf{x}_{i},
\boldsymbol{\eta}_{i}
\right),
\label{eq:h_residual}
\end{equation}

where $H_{0}$ denotes the baseline operator obtained through physical interpolation or the hypsometric transformation, and $\Delta H_{\phi}$ denotes the residual predicted by the neural network. In contrast, ATMS brightness temperatures and GNSS-RO bending angles, which cannot be obtained using simple horizontal and vertical interpolation, are predicted directly by the neural operator.

\subsubsection{Training data and optimization}

The observation operator was trained using GSI diagnostic data from the NOAA UFS GEFS-v13 replay.
The observations include ATMS brightness temperatures from channels 2--22 aboard Suomi-NPP and NOAA-20, as well as temperature, specific humidity, zonal and meridional winds, station pressure, and GNSS-RO bending angles.
Quality-controlled real observations were used as supervised targets.
GSI-simulated observations and observation innovations were retained only for diagnostic purposes and were not provided as inputs to $H_{\phi}$.
Although the ATMS observation operator was trained across channels 2--22, only channels 7--9 were used in the DADA assimilation experiments because they could be represented with sufficient accuracy from the atmospheric state variables available to the forecasting models.
The performance across ATMS channels and the selection of channels 7--9 are analyzed in supplementary material.

Data from 2020 were used for the initial training of the observation operator, while data from 2021 were used for architecture selection, hyperparameter selection, and checkpoint selection. After fixing the model configuration, the final observation operator was retrained using the combined 2020 and 2021 data. Data from 2022 were retained as an independent evaluation set and were not used for training or model selection. The final DADA forecast evaluation was conducted separately using data from 2023.

The atmospheric-state variables, continuous query variables, and observation targets were standardized using variable-specific or observation-group-specific means and standard deviations. For temperature, specific humidity, winds, and station pressure, the residuals between the real observations and their physical baselines were normalized. ATMS brightness temperatures and GNSS-RO bending angles were normalized directly by observation type and channel. The observation operator was trained using the following normalized mean squared error:

\begin{equation}
\mathcal{L}_{H}
=
\frac{1}{N}
\sum_{i=1}^{N}
\left(
\widehat{y}^{\mathrm{norm}}_{i}
-
y^{\mathrm{norm}}_{i}
\right)^{2}.
\label{eq:h_training_loss}
\end{equation}

The model was trained for up to 15 epochs using AdamW with a learning rate of $3 \times 10^{-4}$, a weight decay of $1 \times 10^{-5}$, and a batch size of 32,768. Training used bfloat16 mixed precision, and the global gradient norm was clipped at 5.0. After model selection, the fixed observation operator $H_{\phi}$ was used in all DADA experiments. Its parameters were not updated for individual assimilation cases.

\subsection{Datasets and evaluation metrics}

The observation operator $H_{\phi}$ was trained using ERA5 atmospheric states at $0.25^{\circ}$ resolution and real observations from GSI diagnostics of the NOAA UFS GEFS-v13 replay. Data from 2020 were used for initial training, 2021 for model selection, and the final operator was retrained using the combined 2020--2021 data. Data from 2022 were reserved for independent evaluation.

DADA was evaluated using data from 2023. GFS and HealDA analyses were used as background states, while ERA5 served as the reference for evaluating adapted states and forecasts. We applied DADA to Aurora, FengWu, GraphCast, AIFS, and stochastic FCNv3. The multi-model forecast evaluation used 24 cases initialized at 06 UTC on the first and fifteenth days of each month, with forecast lead times of up to 10 days. Using Aurora with GFS backgrounds, we additionally evaluated the discrepancy of GFS and DADA-adapted states from ERA5 at 6-h intervals throughout 2023.

Deterministic forecasts and adapted states were evaluated using latitude-weighted RMSE against ERA5 for Z500, T850, U500, V500, and Q700. Probabilistic forecasts from stochastic FCNv3 were evaluated using latitude-weighted CRPS computed from 10 ensemble members.

\bibliography{sn-bibliography}

\clearpage

\setcounter{table}{0}
\setcounter{figure}{0}
\setcounter{equation}{0}

\renewcommand{\tablename}{Supplementary Table}
\renewcommand{\figurename}{Supplementary Fig.}
\renewcommand{\thetable}{\arabic{table}}
\renewcommand{\thefigure}{\arabic{figure}}
\renewcommand{\theequation}{S\arabic{equation}}
\renewcommand{\theHtable}{S\arabic{table}}
\renewcommand{\theHfigure}{S\arabic{figure}}
\renewcommand{\theHequation}{S\arabic{equation}}

\section*{Supplementary Information}

\begin{table*}[h]
  \centering
  \caption{Observation-space errors (K) for ATMS brightness temperatures. Centered SD is calculated after removing the channel-wise mean GSI/CRTM error.}
  \label{tab:supp_obs_operator}
  \resizebox{0.5\linewidth}{!}{%
  \begin{tabular}{lccc}
    \toprule
    \multirow{2}{*}{\textbf{Channel}}
    & \multirow{2}{*}{\textbf{$H_{\phi}$ RMSE}}
    & \multicolumn{2}{c}{\textbf{GSI/CRTM}} \\
    \cmidrule(lr){3-4}
    & & Raw RMSE & Centered SD \\
    \midrule
ATMS 2  & \textbf{2.476954} & 3.155379 & 3.152329 \\
ATMS 3  & \textbf{1.958710} & 2.625560 & 2.624673 \\
ATMS 4  & \textbf{1.164067} & 1.959157 & 1.707803 \\
ATMS 5  & \textbf{0.460943} & 1.074024 & 0.595001 \\
ATMS 6  & \textbf{0.192121} & 0.408941 & 0.193317 \\
ATMS 7  & \textbf{0.137100} & 0.607351 & 0.212642 \\
ATMS 8  & 0.173212 & 0.992811 & \textbf{0.140191} \\
ATMS 9  & 0.237705 & 1.009791 & \textbf{0.202866} \\
ATMS 10 & 0.577272 & 0.716508 & \textbf{0.210620} \\
ATMS 11 & 1.347094 & 0.692231 & \textbf{0.220185} \\
ATMS 12 & 2.739497 & 0.793479 & \textbf{0.220285} \\
ATMS 13 & 4.331110 & 1.065720 & \textbf{0.311399} \\
ATMS 14 & 5.365727 & 1.181464 & \textbf{0.450714} \\
ATMS 15 & 5.810463 & 1.011856 & \textbf{0.699220} \\
ATMS 16 & \textbf{3.129026} & 3.432366 & 3.428995 \\
ATMS 17 & \textbf{1.964668} & 2.423137 & 2.339071 \\
ATMS 18 & 1.276355 & 1.653800 & \textbf{1.229600} \\
ATMS 19 & 1.383326 & 1.398924 & \textbf{1.170310} \\
ATMS 20 & 1.522965 & 1.392776 & \textbf{1.199080} \\
ATMS 21 & 1.686363 & 1.348542 & \textbf{1.252914} \\
ATMS 22 & 1.782852 & 1.314962 & \textbf{1.314958} \\
    \bottomrule
  \end{tabular}%
  }
\end{table*}

\begin{figure}[t]
    \centering
    \includegraphics[width=\columnwidth]{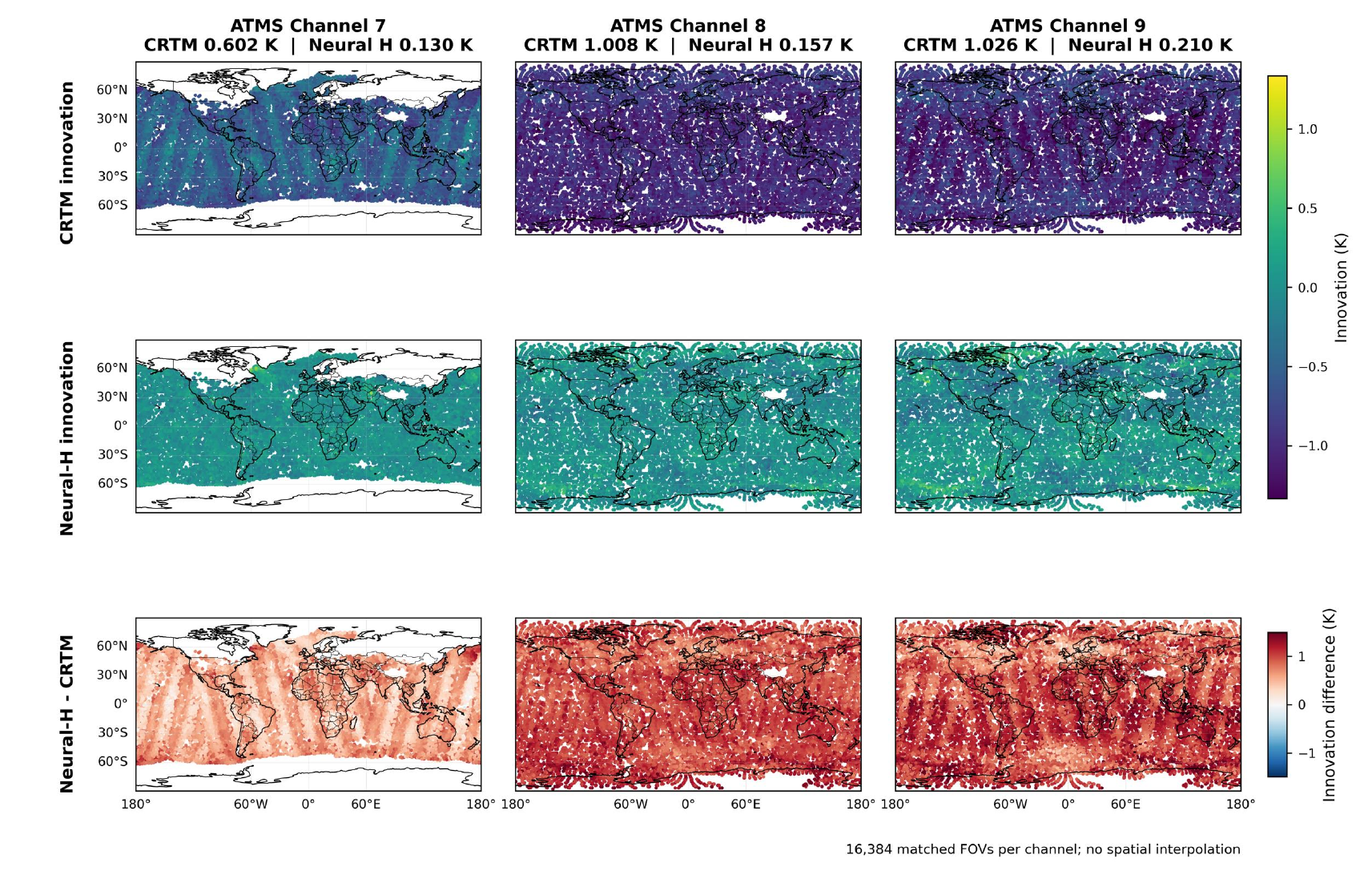}
    \caption{Full-day comparison of ATMS innovations produced by GSI/CRTM and the learned observation operator on 30 November 2023. Observations from the 00, 06, 12, and 18~UTC cycles were combined, and the columns show ATMS channels 7, 8, and 9 used in DADA. The top row shows the GSI/CRTM innovation, $d_{\mathrm{CRTM}} = y - H_{\mathrm{CRTM}}(x)$; the middle row shows the innovation obtained using the learned observation operator, $d_{\phi} = y - H_{\phi}(x)$; and the bottom row shows their signed difference, $d_{\phi} - d_{\mathrm{CRTM}}$. The RMSE values for GSI/CRTM and $H_{\phi}$ are reported above each column. Each channel contains 16,384 matched observation footprints. Only observed field-of-view locations are displayed, without spatial interpolation.}
    \label{fig:sup_inno}
\end{figure}

\begin{table*}[t!]
  \centering
  \caption{
    Forecast RMSE of Aurora under different background states and DADA supervision targets. Units are indicated for each variable, and lower values denote better performance. The best and second-best values for
    each variable and forecast lead time are shown in bold and underlined, respectively. 
  }
  \label{tab:aurora_supervision}

  \setlength{\tabcolsep}{5pt}
  \renewcommand{\arraystretch}{1.0}
  \resizebox{!}{0.4\textheight}{%
  \begin{tabular}{@{}lcccc@{}}
    \toprule
    & \multicolumn{4}{c}{\textbf{Forecast lead time}} \\
    \cmidrule(lr){2-5}
    \textbf{Initialization} & \textbf{Day 1} & \textbf{Day 3} & \textbf{Day 7} & \textbf{Day 9} \\
    \midrule

    \multicolumn{5}{l}{\textit{Z500 (gpm)}} \\
    ERA5
      & 5.95
      & 17.70
      & 61.33
      & \underline{79.10} \\
    GFS
      & 12.17
      & 24.92
      & 66.21
      & 84.43 \\
    GFS + DADA (observations)
      & 6.62
      & 19.37
      & 62.59
      & 81.95 \\
    GFS + DADA (ERA5 state)
      & \underline{5.21}
      & \underline{16.39}
      & \underline{60.30}
      & 80.40 \\
    HealDA
      & 10.36
      & 23.14
      & 67.08
      & 86.00 \\
    HealDA + DADA (observations)
      & 5.30
      & 18.11
      & 60.69
      & 80.60 \\
    HealDA + DADA (ERA5 state)
      & \textbf{4.91}
      & \textbf{15.30}
      & \textbf{59.21}
      & \textbf{77.29} \\

    \addlinespace[2pt]
    \multicolumn{5}{l}{\textit{T850 (K)}} \\
    ERA5
      & \underline{0.680}
      & 1.320
      & 3.011
      & 3.697 \\
    GFS
      & 1.104
      & 1.631
      & 3.149
      & 3.801 \\
    GFS + DADA (observations)
      & 0.827
      & 1.414
      & 3.033
      & 3.657 \\
    GFS + DADA (ERA5 state)$^{*}$
      & \textbf{0.651}
      & \underline{1.197}
      & \textbf{2.922}
      & \underline{3.588} \\
    HealDA
      & 0.971
      & 1.580
      & 3.298
      & 3.877 \\
    HealDA + DADA (observations)
      & 0.776
      & 1.363
      & 3.052
      & 3.728 \\
    HealDA + DADA (ERA5 state)$^{*}$
      & 0.718
      & \textbf{1.151}
      & \underline{2.977}
      & \textbf{3.575} \\

    \addlinespace[2pt]
    \multicolumn{5}{l}{\textit{U500 (m\,s$^{-1}$)}} \\
    ERA5
      & \underline{1.682}
      & 3.346
      & 7.348
      & \underline{8.990} \\
    GFS
      & 2.270
      & 4.025
      & 7.829
      & 9.263 \\
    GFS + DADA (observations)
      & 1.836
      & 3.524
      & 7.489
      & 9.225 \\
    GFS + DADA (ERA5 state)$^{*}$
      & \textbf{1.444}
      & \underline{2.982}
      & \underline{7.215}
      & 9.050 \\
    HealDA
      & 2.282
      & 3.914
      & 7.955
      & 9.507 \\
    HealDA + DADA (observations)
      & 1.821
      & 3.478
      & 7.348
      & 9.115 \\
    HealDA + DADA (ERA5 state)$^{*}$
      & 1.687
      & \textbf{2.938}
      & \textbf{7.169}
      & \textbf{8.741} \\

    \addlinespace[2pt]
    \multicolumn{5}{l}{\textit{V500 (m\,s$^{-1}$)}} \\
    ERA5
      & \underline{1.685}
      & 3.404
      & 7.716
      & \underline{9.175} \\
    GFS
      & 2.239
      & 4.124
      & 8.011
      & 9.677 \\
    GFS + DADA (observations)
      & 1.837
      & 3.562
      & 7.832
      & 9.432 \\
    GFS + DADA (ERA5 state)$^{*}$
      & \textbf{1.445}
      & \underline{3.015}
      & \textbf{7.544}
      & 9.253 \\
    HealDA
      & 2.254
      & 4.024
      & 8.364
      & 9.694 \\
    HealDA + DADA (observations)
      & 1.841
      & 3.544
      & 7.837
      & 9.376 \\
    HealDA + DADA (ERA5 state)$^{*}$
      & 1.706
      & \textbf{2.993}
      & \underline{7.645}
      & \textbf{8.991} \\

    \addlinespace[2pt]
    \multicolumn{5}{l}{\textit{Q700 (g\,kg$^{-1}$)}} \\
    ERA5
      & \underline{0.689}
      & 1.119
      & 1.727
      & 1.908 \\
    GFS
      & 0.875
      & 1.225
      & 1.769
      & 1.927 \\
    GFS + DADA (observations)
      & 0.778
      & 1.169
      & 1.744
      & 1.939 \\
    GFS + DADA (ERA5 state)$^{*}$
      & \textbf{0.612}
      & \underline{0.989}
      & \textbf{1.681}
      & \underline{1.903} \\
    HealDA
      & 0.831
      & 1.208
      & 1.802
      & 1.958 \\
    HealDA + DADA (observations)
      & 0.754
      & 1.157
      & 1.749
      & 1.943 \\
    HealDA + DADA (ERA5 state)$^{*}$
      & 0.698
      & \textbf{0.977}
      & \underline{1.706}
      & \textbf{1.863} \\
    \bottomrule
  \end{tabular}%
  }
\end{table*}

\begin{table*}[t!]
\centering
\small
\caption{Forecast RMSE of Aurora initialized from ERA5, GFS, or
GFS adapted by DADA using observational constraints at $T$ only
or at both $T$ and $T+6$.
Forecast lead times are measured from the analysis time $T$.
Units are indicated for each variable, and lower values indicate
better performance. The lowest RMSE for each variable and forecast
lead time is shown in bold.}
\label{tab:da_window_ablation}
\begin{tabular}{lcccc}
\toprule
Initialization & Day 1 & Day 3 & Day 7 & Day 9 \\
\midrule

\multicolumn{5}{l}{Z500 (gpm)} \\
ERA5
& \textbf{5.95} & \textbf{17.70}
& \textbf{61.33} & \textbf{79.10} \\
GFS
& 12.17 & 24.92 & 66.21 & 84.43 \\
GFS + DADA ($T$ only)
& 9.16 & 22.39 & 65.17 & 85.21 \\
GFS + DADA ($T{:}T+6$)
& 6.62 & 19.39 & 62.48 & 81.82 \\
\midrule

\multicolumn{5}{l}{T850 (K)} \\
ERA5
& \textbf{0.680} & \textbf{1.320}
& \textbf{3.011} & 3.697 \\
GFS
& 1.104 & 1.631 & 3.149 & 3.801 \\
GFS + DADA ($T$ only)
& 0.954 & 1.541 & 3.106 & 3.793 \\
GFS + DADA ($T{:}T+6$)
& 0.827 & 1.415 & 3.035 & \textbf{3.658} \\
\midrule

\multicolumn{5}{l}{U500 (m s$^{-1}$)} \\
ERA5
& \textbf{1.682} & \textbf{3.346}
& \textbf{7.348} & \textbf{8.990} \\
GFS
& 2.270 & 4.025 & 7.829 & 9.263 \\
GFS + DADA ($T$ only)
& 2.117 & 3.807 & 7.705 & 9.280 \\
GFS + DADA ($T{:}T+6$)
& 1.836 & 3.526 & 7.486 & 9.228 \\
\midrule

\multicolumn{5}{l}{V500 (m s$^{-1}$)} \\
ERA5
& \textbf{1.685} & \textbf{3.404}
& \textbf{7.716} & \textbf{9.175} \\
GFS
& 2.239 & 4.124 & 8.011 & 9.677 \\
GFS + DADA ($T$ only)
& 2.098 & 3.894 & 7.917 & 9.654 \\
GFS + DADA ($T{:}T+6$)
& 1.837 & 3.563 & 7.834 & 9.454 \\
\midrule

\multicolumn{5}{l}{Q700 (g kg$^{-1}$)} \\
ERA5
& \textbf{0.689} & \textbf{1.119}
& \textbf{1.727} & \textbf{1.908} \\
GFS
& 0.875 & 1.225 & 1.769 & 1.927 \\
GFS + DADA ($T$ only)
& 0.850 & 1.210 & 1.752 & 1.934 \\
GFS + DADA ($T{:}T+6$)
& 0.778 & 1.169 & 1.745 & 1.938 \\
\bottomrule
\end{tabular}
\end{table*}

\subsection*{Observation operator performance}
The learned observation operator $H_{\phi}$ is model-specific because AI weather forecasting models differ in the atmospheric variables, vertical levels, and preprocessing conventions exposed by their state representations.
Accordingly, DADA uses a separate $H_{\phi}$ for each target forecasting model, constructed from the atmospheric state variables available from that model.
Here, we analyze the FCNv3-compatible $H_{\phi}$ as a representative case.

Conventional physics-based observation operators such as CRTM~\cite{johnson2023community} are designed to exploit the relatively rich atmospheric state and ancillary information available in numerical weather prediction systems.
In contrast, the FCNv3-compatible $H_{\phi}$ is restricted to the state representation exposed by FCNv3, which in this study includes atmospheric variables on 13 pressure levels together with a limited set of surface variables.
The purpose of $H_{\phi}$ is therefore not to replace CRTM as a general-purpose radiative transfer model, but to provide a differentiable observation-space interface that is directly compatible with the state representation of each pretrained AI forecasting model.

We evaluated the FCNv3-compatible $H_{\phi}$ using ATMS brightness temperatures on an independent test dataset.
Supplementary Table~\ref{tab:supp_obs_operator} reports the RMSE of $H_{\phi}$ for ATMS channels 2--22 together with the corresponding GSI/CRTM errors.
The GSI/CRTM values were calculated from the bias-uncorrected observation equivalents stored in the GSI diagnostic files of the NOAA UFS GEFS-v13 replay.
Because the raw GSI/CRTM RMSE contains substantial channel-dependent mean biases, we additionally report the centered standard deviation after removing the channel-wise mean error.
These values are provided as a diagnostic reference and do not represent the operational variational bias correction used by GSI.

The performance of $H_{\phi}$ varied substantially across ATMS channels.
Channels 7--9 were predicted accurately, with RMSEs of 0.137, 0.173, and 0.238 K, respectively, and were therefore the only ATMS radiance channels used in the DADA experiments.
In contrast, the errors of $H_{\phi}$ increased substantially for channels 10--15 relative to GSI/CRTM.
For channels 11--15, the RMSE of $H_{\phi}$ increased from approximately 1.35 K to 5.81 K, whereas the centered GSI/CRTM scatter remained approximately 0.22--0.70 K.
We therefore excluded channels 10--15 from DADA assimilation rather than using radiance observations that could not be represented with sufficient accuracy from the available FCNv3 state.

Supplementary Fig.~\ref{fig:sup_inno} provides a full-day spatial comparison of ATMS innovations for channels 7--9 used in DADA, following the spatial visualization of observation-operator errors in prior work~\cite{lieberman2026deep}.
The GSI/CRTM and learned-operator innovations are evaluated at the same observation footprints, showing that $H_{\phi}$ produces small residual errors with coherent spatial structure for the selected channels.

These results should not be interpreted as showing that the learned operator is generally more accurate or more complete than CRTM.
Rather, they show that, despite operating on a substantially more restricted AI-model state, a model-compatible learned observation operator can provide sufficiently accurate differentiable observation constraints for selected channels.

\subsection*{Role of observation-space supervision}

To assess whether observation-space supervision can provide an effective alternative to direct access to ERA5 states, we compared the default DADA configuration with a configuration that bypasses the learned observation operator $H_{\phi}$ and directly uses 1\% of the ERA5 grid points as state-space optimization targets.
Rather than using the complete ERA5 fields, we retained only 1\% of the grid points to approximate a similarly sparse supervision regime.
This ERA5-supervised configuration is not intended as an operational setting, since ERA5 is unavailable in real time, but serves as a reference for assessing how much of the benefit of direct state-space supervision can be recovered from observations.

As shown in Table~\ref{tab:aurora_supervision}, DADA with direct ERA5-state supervision achieved the lowest RMSE for most variables and forecast lead times.
Two factors are consistent with this improvement.
First, because Aurora was trained on ERA5, directly supervising the atmospheric state with ERA5 reduces the distributional mismatch between the GFS or HealDA background and the model's training distribution.
In addition, because forecast performance is evaluated against ERA5, this configuration benefits from direct alignment between the optimization target and the verification reference.

Second, ERA5-state supervision constrains more than the atmospheric state at the analysis time.
Forecasts initialized directly from ERA5 use only the state at the analysis time, whereas DADA is optimized over an assimilation window that also constrains the state six hours later.
The optimization therefore affects both the adapted initial state and the short-range trajectory generated by the frozen Aurora model.
The fact that ERA5-supervised DADA outperformed direct ERA5 initialization for some variables and lead times is consistent with an additional contribution from the trajectory constraint, beyond alignment with the ERA5 training distribution.

DADA using $H_{\phi}$ generally produced higher errors than direct ERA5-state supervision, but achieved forecast performance comparable to ERA5 initialization across multiple variables and lead times.
This result shows that, even without direct access to ERA5 states, observation-space constraints through $H_{\phi}$ recover a substantial fraction of the improvement obtained from direct state-space supervision.

Taken together, these results are consistent with two complementary effects of DADA: adapting the background toward an initial state that is more compatible with the training distribution of the forecasting model, and constraining the subsequent short-range trajectory with observations.
The role of $H_{\phi}$ is therefore not to reconstruct or replace ERA5, but to provide a differentiable interface through which these two effects can be induced from real-world observations when ERA5 states are unavailable in real time.

\subsection*{Sensitivity to assimilation-window length}

To examine how the observation window affects DADA performance,
we compared a setting using real observations at $T$ only with
a setting using observations at both $T$ and $T+6$.
Both settings were evaluated at the same verification times and
model integration lengths.
The $T{:}T+6$ setting was used in the main experiments.

As shown in Table~\ref{tab:da_window_ablation}, using observations
at $T$ only reduced errors relative to the GFS control for most
variables and forecast lead times.
The improvements therefore cannot be attributed solely to adding
observations at $T+6$.
Including observations at $T+6$ generally reduced errors further,
with larger additional improvements at short forecast lead times.

Together with Supplementary Table~\ref{tab:aurora_supervision},
these results support the interpretation that adjusting the
background to the target forecasting model contributes to the
forecast improvements.
Observations at $T$ constrain the analysis generated by the model,
whereas observations at $T+6$ also constrain the subsequent
forecast state.
However, adding the $T+6$ constraint also provides later
observational information, so this comparison cannot determine
the separate contributions of these effects.

The time span used in the main experiments is also used in
established data assimilation systems.
DADA starts from the background at $T-6$ and includes model
predictions through $T+6$ in the optimization, covering 12~h.
This is the same duration as the assimilation window used in
ECMWF's IFS long-window 4D-Var configuration.
Although the observation timings and procedures differ, both
approaches use observations to constrain atmospheric evolution
over this temporal scale.

Observational constraints were applied at 6-h intervals because
the forecasting models used here provide outputs at this frequency.
The $T{:}T+6$ setting therefore constrains the analysis and the
next forecast state together.
Models with more frequent outputs would allow observational
constraints to be applied at shorter intervals.
Intermediate windows ending between $T$ and $T+6$ were not
evaluated in this study.

\subsection*{Power spectrum analysis of DADA analyses}

To examine whether DADA excessively smooths small-scale spatial structures in the analysis fields, we compared the spherical-harmonic power spectra of ERA5, the Aurora GFS control, and the DADA analyses. The analysis was conducted using 24 cases initialized at 06 UTC on the first day of each month in 2023. After normalizing the spectrum by the total power over all nonzero spherical-harmonic degrees, we defined high-wavenumber energy as the sum of the normalized power at total degrees $\ell \geq 100$. Table~\ref{tab:power_spectrum_hf} reports the high-wavenumber energy relative to ERA5.

The GFS control exhibited lower high-wavenumber energy than ERA5 for all evaluated variables. In contrast, the high-wavenumber energy ratios of the DADA analyses ranged from 1.003 to 1.581, indicating levels comparable to or greater than those of ERA5. Q700 showed nearly the same high-wavenumber energy as ERA5, whereas T850, Z500, U500, and V500 exhibited higher values. These results indicate that DADA does not reduce forecast errors by removing small-scale spatial variability or excessively smoothing the analysis fields.

\begin{figure}[t]
    \centering
    \includegraphics[width=\columnwidth]{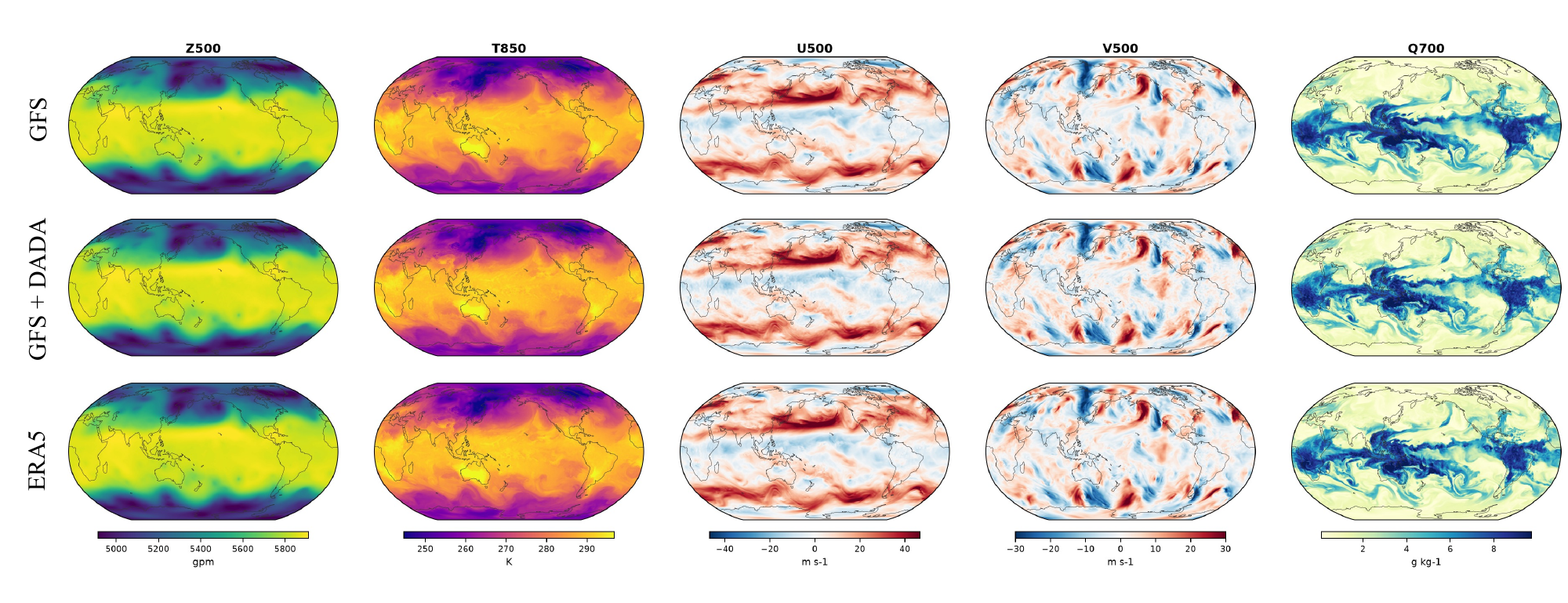}
    \caption{Qualitative comparison of the GFS, GFS+DADA, and ERA5 analysis fields. Rows show GFS, GFS adapted by DADA, and ERA5 from top to bottom, while columns show Z500, T850, U500, V500, and Q700. The DADA analyses preserve small-scale spatial structures without apparent excessive smoothing while reducing the errors of the GFS background.}
    \label{fig:sup_fig1}
\end{figure}

As also observed qualitatively in Fig.~\ref{fig:sup_fig1}, the DADA analyses preserve small-scale spatial structures without excessive smoothing. In some regions, these structures are more pronounced than those in the GFS control.

Together, these quantitative and qualitative results show that the lower analysis RMSE achieved by DADA is not obtained through excessive smoothing or the removal of small-scale spatial variability.

\subsection*{Visualization of Spatial Forecast Errors}

\begin{figure}[t]
    \centering
    \includegraphics[width=\columnwidth]{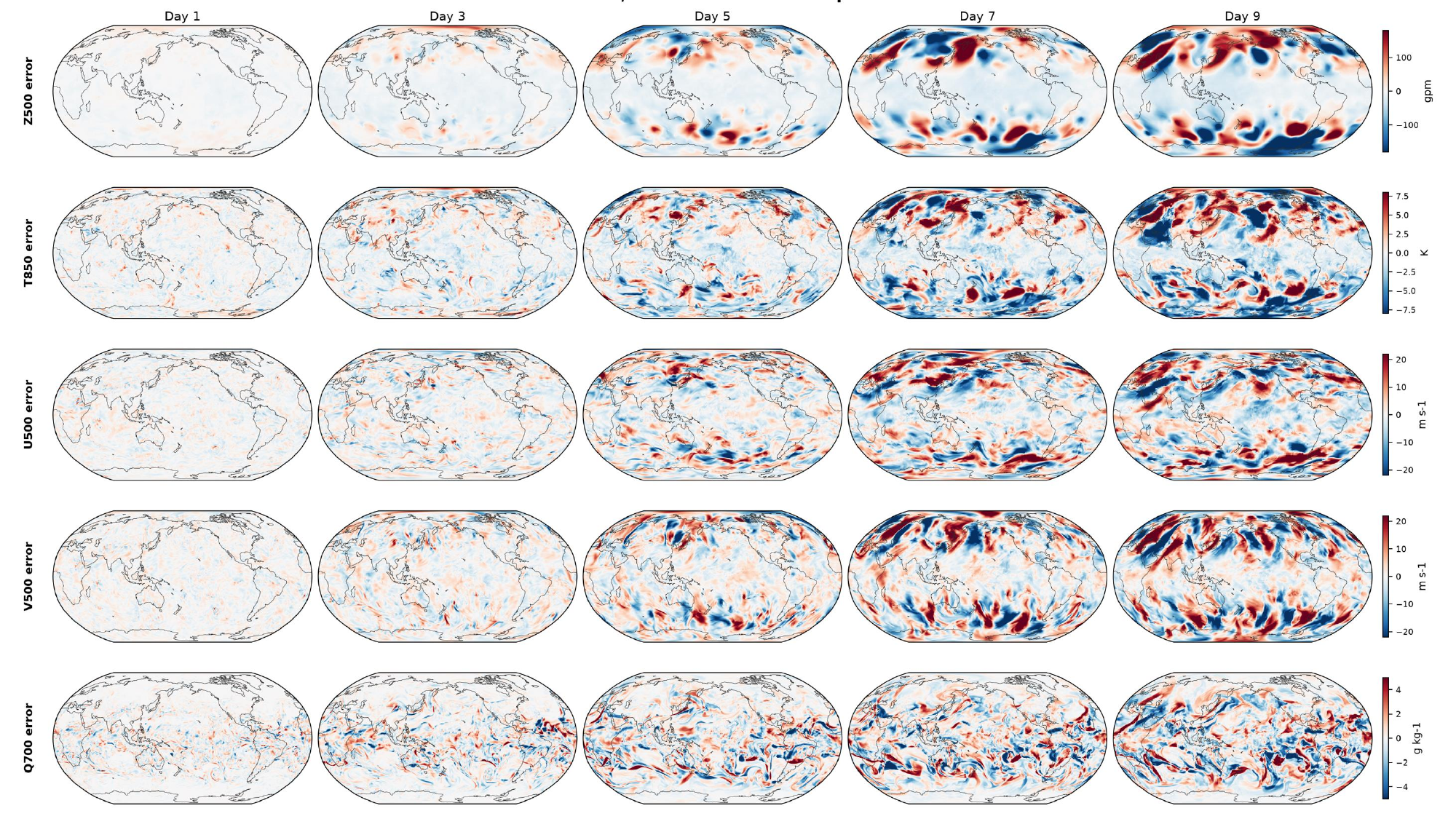}
    \caption{Spatial forecast errors of DADA against ERA5 for the March 2023 case. Columns show forecast lead times of 1, 3, 5, 7, and 9 days, and rows show errors for Z500, T850, U500, V500, and Q700. Forecast errors generally increase with lead time, with relatively large errors concentrated in the high-latitude regions.}
    \label{fig:sup_fig2}
\end{figure}

\begin{figure}[t]
    \centering
    \includegraphics[width=\columnwidth]{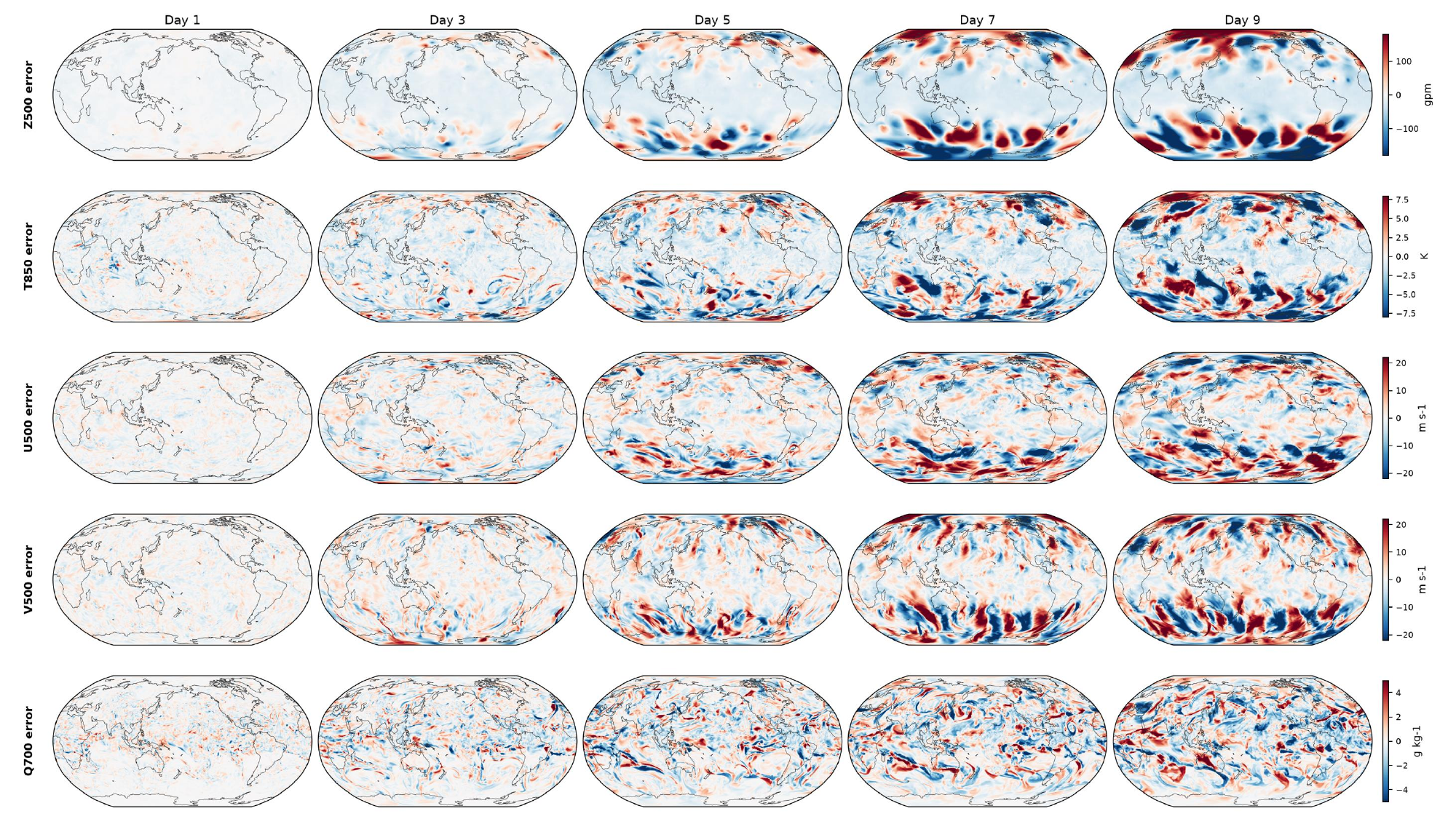}
    \caption{Spatial forecast errors of DADA against ERA5 for the June 2023 case. Columns show forecast lead times of 1, 3, 5, 7, and 9 days, and rows show errors for Z500, T850, U500, V500, and Q700. Forecast errors generally increase with lead time, with relatively large errors concentrated in the high-latitude regions.}
    \label{fig:sup_fig3}
\end{figure}

\begin{figure}[t]
    \centering
    \includegraphics[width=\columnwidth]{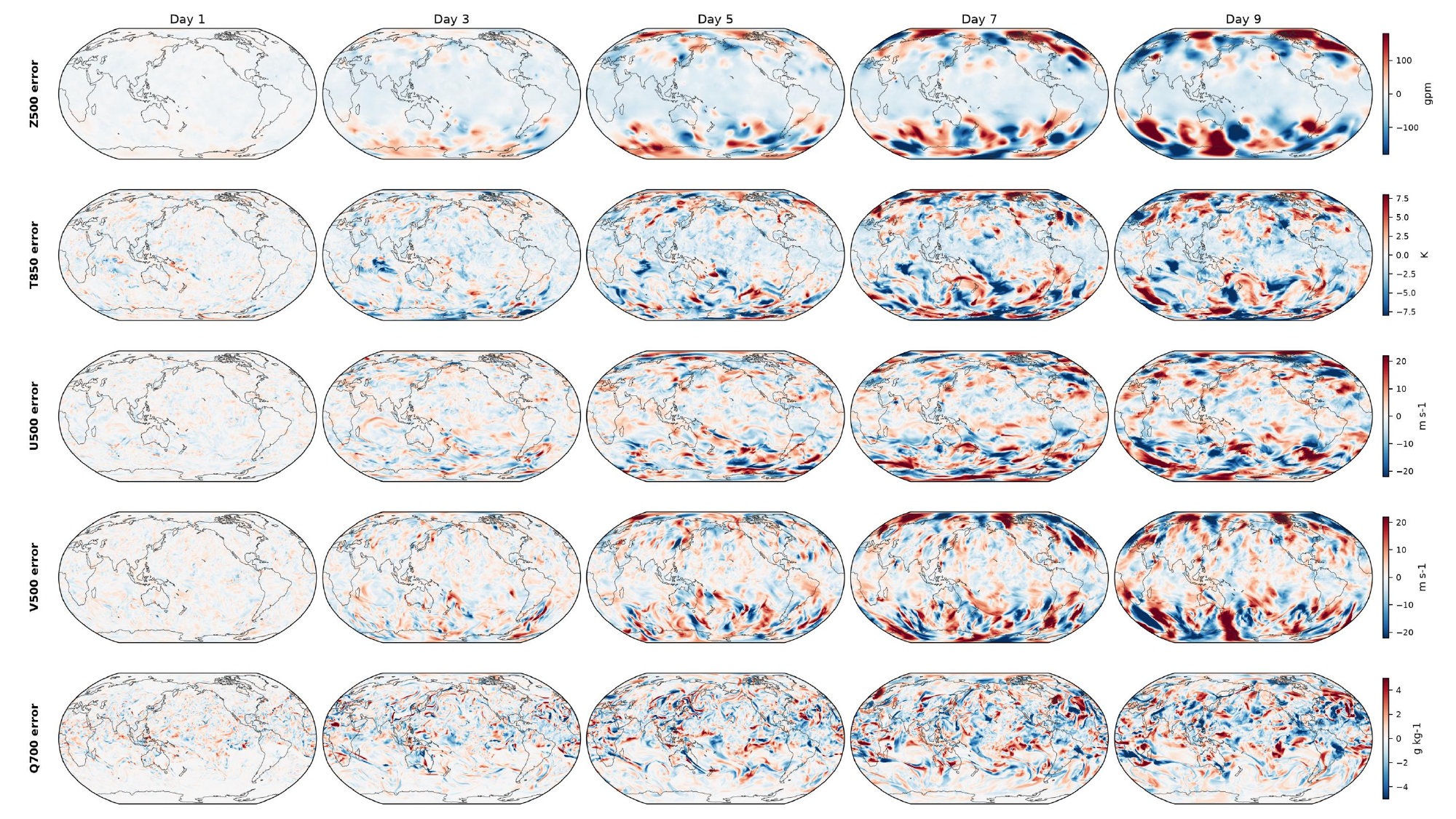}
    \caption{Spatial forecast errors of DADA against ERA5 for the September 2023 case. Columns show forecast lead times of 1, 3, 5, 7, and 9 days, and rows show errors for Z500, T850, U500, V500, and Q700. Forecast errors generally increase with lead time, with relatively large errors concentrated in the high-latitude regions.}
    \label{fig:sup_fig4}
\end{figure}

\begin{figure}[t]
    \centering
    \includegraphics[width=\columnwidth]{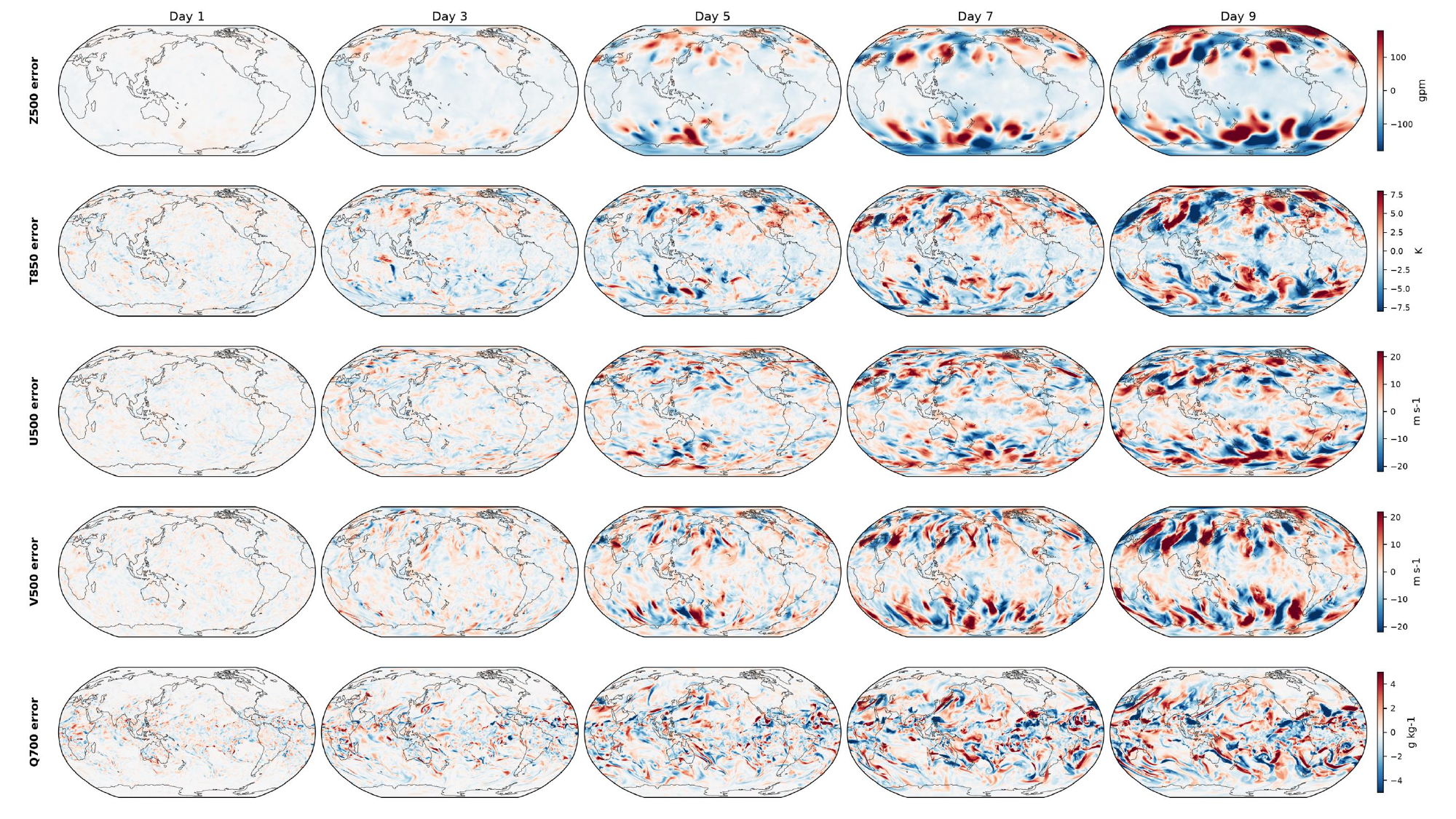}
    \caption{Spatial forecast errors of DADA against ERA5 for the December 2023 case. Columns show forecast lead times of 1, 3, 5, 7, and 9 days, and rows show errors for Z500, T850, U500, V500, and Q700. Forecast errors generally increase with lead time, with relatively large errors concentrated in the high-latitude regions.}
    \label{fig:sup_fig5}
\end{figure}

To examine the spatial characteristics of the forecast errors, we visualized the error fields of DADA forecasts against ERA5 at lead times of 1, 3, 5, 7, and 9 days. Figures~~\ref{fig:sup_fig2}-~\ref{fig:sup_fig5} show representative cases from March, June, September, and December, respectively, covering different seasons of 2023.

Across the four cases, forecast errors generally increased with lead time and exhibited clear spatial structures. Relatively large errors were concentrated in the high-latitude regions, particularly for Z500, U500, and V500, while errors over the low- and mid-latitude regions were generally smaller. The high-latitude error patterns became increasingly pronounced at longer forecast lead times. T850 showed a similar tendency, although with smaller-scale spatial variations, while Q700 exhibited more spatially distributed and fine-scale error structures.

Despite seasonal differences in the detailed spatial patterns, the overall characteristics were similar across March, June, September, and December. These results indicate that the forecast errors of DADA are not uniformly distributed over the globe, but are predominantly associated with high-latitude regions, particularly at longer forecast lead times.

\begin{table*}[t]
\centering
\caption{High-wavenumber energy ratios of the DADA analyses. The power spectrum was normalized by the total power across all nonzero spherical-harmonic degrees, and high-wavenumber energy was defined as the sum of the normalized power over total degrees $\ell \geq 100$. Values are reported as the mean and standard deviation over 24 cases initialized at 06 UTC on the first and fifteenth days of each month in 2023. A ratio of 1 indicates the same high-wavenumber energy as ERA5, whereas values below or above 1 indicate lower or higher high-wavenumber energy than ERA5, respectively.}
\label{tab:power_spectrum_hf}
\begin{tabular}{lccc}
\toprule
Variable
& GFS control / ERA5
& DADA / ERA5
& DADA min--max \\
\midrule

Z500
& $0.936 \pm 0.136$
& $1.581 \pm 0.087$
& $1.458$--$1.758$ \\

T850
& $0.745 \pm 0.046$
& $1.088 \pm 0.038$
& $1.018$--$1.140$ \\

U500
& $0.883 \pm 0.037$
& $1.446 \pm 0.053$
& $1.340$--$1.553$ \\

V500
& $0.837 \pm 0.042$
& $1.447 \pm 0.048$
& $1.358$--$1.512$ \\

Q700
& $0.833 \pm 0.038$
& $1.003 \pm 0.025$
& $0.973$--$1.046$ \\

\bottomrule
\end{tabular}
\end{table*}

\end{document}